\documentclass[12pt]{article}

\usepackage{amsmath}
\usepackage{amsfonts}
\usepackage{amscd}
\usepackage{amsthm}
\usepackage{setspace}
\usepackage{graphicx}
\usepackage{authblk}
\usepackage{caption}

\def\Z{\mathbb{Z}}

\def\R{\mathbb{R}}
\def\C{\mathbb{C}}
\def\P{\mathbb{P}}

\def\til{\tilde}

\begin{document}

\begin{titlepage}

\begin{flushright}
YITP-15-104
\end{flushright}

\vskip 1cm

\begin{center}

{\large Gauge Groups and Matter Fields on Some Models of \\
F-theory without Section}

\vskip 1.2cm

Yusuke Kimura$^1$
\vskip 0.4cm
{\it $^1$Yukawa Institute for Theoretical Physics, Kyoto University, Kyoto 606-8502, Japan}
\vskip 0.4cm
E-mail: kimura@yukawa.kyoto-u.ac.jp

\vskip 1.5cm
\abstract{We investigate F-theory on an elliptic Calabi-Yau 4-fold without a section to the fibration. To construct an elliptic Calabi-Yau 4-fold without a section, we introduce families of elliptic K3 surfaces which do not admit a section. A product K3 $\times$ K3, with one of the K3's chosen from these families of elliptic K3 surfaces without a section, realises an elliptic Calabi-Yau 4-fold without a section. We then compactify F-theory on such K3 $\times$ K3's.
\par We determine the gauge groups and matter fields which arise on 7-branes for these models of F-theory compactifications without a section. Since each K3 $\times$ K3 constructed does not have a section, gauge groups arising on 7-branes for F-theory models on constructed K3 $\times$ K3's do not have $U(1)$-part. Interestingly, exceptional gauge group $E_6$ appears for some cases.}  

\end{center}
\end{titlepage}

\tableofcontents

\section{Introduction}
\par F-theory\cite{Vaf, MV1, MV2} is a non-perturbative extension of type IIB superstring theory, which is considered to encompass strong coupling regime. Description of high energy regime requires that a coupling constant vary over a base 3-fold. 
This is realised in F-theory, where the theory is compactified on an elliptic Calabi-Yau 4-fold over a base 3-fold, with fiber being an elliptic curve. A complex structure of a fiber as a parameter 
is identified with a coupling constant. So, in the framework of F-theory the coupling constant becomes a function over a base 3-fold. $SL(2,\Z)$-symmetry present on an elliptic fiber can be seen as a geometrical realisation of S-duality. 
\par Many aspects of gauge theory are described in the geometrical language in F-theory. {\it The discriminant locus} is the set of points in the base, over which an elliptic fiber becomes singular (i.e. not a smooth elliptic curve). The discriminant locus signals locations of 7-branes. 7-branes are wrapped on irreducible components of the discriminant locus. Not only the location of 7-branes, but the information on the structures of gauge groups can be read by studying the discriminant locus. Singular fibers lying over the discriminant locus encode the information on gauge groups in low energy theory. A type of singular fiber specifies the gauge groups which arise on 7-branes wrapped on an irreducible component of discriminant locus. So, by looking at a type of a singular fiber, the structure of gauge groups can be known. The types of singular fibers of an elliptic surface were classified by Kodaira\cite{Kod}.  
\par Many models of F-theory on elliptic 4-folds have been studied, mostly 4-folds with sections\cite{DW, FI, CGH, GW, KMSS, GKPW, CKP, MPW, BMPW, CGKP, BGK}. However, among all the elliptic Calabi-Yau 4-folds, ones which admit a section to the fibration only form a special subset. F-theory compactifications on elliptic manifolds without a section was considered in \cite{BEFNQ}. Recently, Braun and Morrison\cite{BM} investigated F-theory on spaces which admit elliptic fibration without a section, and Morrison and Taylor\cite{MT section} studied F-theory on elliptic Calabi-Yau manifolds
which do not have a section. For recent models of F-theory without a section, see also \cite{AGGK, KMOPR, GGK, MPTW, MPTW2, CDKPP, LMTW}. 
\par In the present paper, we explore another model of F-theory on elliptic Calabi-Yau 4-folds which do not admit a section. To build a model without section, we consider an elliptic K3 surface times a K3 surface. We choose a direct product K3 $\times$ K3 as an elliptic Calabi-Yau 4-fold, for the reason that the techniques from algebraic geometry enable to perform a detailed analysis of F-theory compactifications on K3 $\times$ K3. 
\par By using the method of algebraic geometry, we construct families of elliptic K3 surfaces without a section to the fibration. By considering a product of an elliptic K3 surface without a section times a K3 surface, we obtain an elliptic Calabi-Yau 4-fold without a section. We explicitly determine gauge groups which arise on 7-branes for F-theory compactified on these spaces. Non-Abelian part of the gauge groups are read from the types of singular fibers of an elliptic K3 surface. For F-theory on K3 $\times$ K3, the number of $U(1)$-factors in the gauge group equals the rank of the Mordell-Weil group of an elliptic K3 surface\cite{MV2}. An elliptic K3 surface we consider in this paper does not have a section. So, gauge groups for these models without a section do not have $U(1)$-part. Interestingly, exceptional gauge group $E_6$ arises in some of our models. 
\par We next investigate matter fields which arise on 7-branes for the models. Matter fields in the F-theory context were studied by Katz and Vafa\cite{KV}. As discussed in \cite{KV}, a singularity of a compactifying space encodes information on charged matters on 7-branes. As 7-branes localised at the singularity move apart from one another, they generate light matter fields, and from geometric viewpoint this corresponds to a deformation of singularity. A singular fiber of ADE-type of an elliptic fibration corresponds to a singularity of ADE-type. When 7-branes move so a singularity of ADE-type deforms accordingly, the corresponding gauge group $G$ breaks to $H$ of one rank less. Resultantly, matters arise on 7-branes in irreducible representations of $H$. Therefore, matter fields on 7-branes can be read by deforming singularities of ADE-type. See for instance \cite{BIKMSV, KMP, Pha} for related issues. For recent advances on the correspondence between matter and geometry, see \cite{MT matter}. Since we limit the consideration to F-theory compactifications on K3 $\times$ K3, all the 7-branes are parallel. So the Yukawa-type interaction is absent for our models of F-theory on K3 $\times$ K3.    
\par The paper is structured as follows: in section 2 we introduce families of elliptic K3 surfaces which do not admit a section. K3 $\times$ K3, with one of the K3's chosen from this introduced class of elliptic K3's, gives an elliptic Calabi-Yau 4-fold without a section. In section 3, we determine gauge groups arising on 7-branes for F-theory compactified on K3 $\times$ K3, with one of the K3's chosen from the class introduced in section 2. By construction, such K3 $\times$ K3 does not admit a section. So, the gauge groups which arise on 7-branes do not have $U(1)$ factor. We also see that for some cases, consideration on monodromy uniquely determines the gauge groups. F-theory on K3 $\times$ K3 gives 4d theory with enhanced $N=2$ supersymmetry. Therefore, the theory is highly constrained by the anomaly cancellation conditions. We check if our solutions are consistent with these anomaly cancellation conditions. In section 4, we calculate the matter spectra which arise on 7-branes, with and without flux. Since potential matter may actually vanish due to anomalies, we can fully determine the matter spectra for some limited cases. In section 5, we derive conclusions on F-theory compactifications on K3 $\times$ K3 without a section.

\section{Class of Elliptic K3 surfaces which do not admit Section}
In this section, we introduce a class of elliptic K3 surfaces which do not admit a section. 
\par We can explicitly construct a class of elliptic K3 surfaces without a section as follows: consider a smooth hypersurface $S$ of bidegree (2,3) in $\P^1\times \P^2$. Then a hypersurface $S$ is a K3 surface.  
By restricting projections from $\P^1\times \P^2$ onto $\P^1$ and $\P^2$ respectively to this hypersurface $S$, we obtain a map onto $\P^1$, and a degree 3 map into $\P^2$. These maps give an explicit description of the hypersurface $S$ as an elliptic fibration over a base $\P^1$. 
$$
\begin{CD}
\P^1\times\P^2 \supset S @>{\rm degree \hskip 1mm 3}>> \mathbb{E}\subset \P^2 \\
@VVV \\
\P^1
\end{CD}
$$ 
\par For a generic hypersurface $S$ described above, pullbacks of the hyperplane classes in $\P^1$ and $\P^2$ generate N\'eron-Severi lattice of $S$, and Picard number $\rho(S)=2$ \cite{RS, Ott}. Let $D_1$ and $D_2$ denote divisors of hypersurface $S$ corresponding to pullbacks of hyperplanes in $\P^1$ and $\P^2$ respectively. Let $p_1$ and $p_2$ denote projections from $\P^1\times \P^2$ onto $\P^1$ and $\P^2$ respectively: $$
\begin{CD}
\P^1\times\P^2 @>{p_2}>> \P^2 \\
@V{p_1}VV \\
\P^1
\end{CD}
$$ Then $p_1^*{\cal O}_{\P^1}(1)=\{pt\}\times \P^2$ and $p_2^*{\cal O}_{\P^2}(1)=\P^1\times line$. Restrictions of these to a hypersurface $S$ are $D_1$ and $D_2$, and $S\sim p_1^*{\cal O}_{\P^1}(2)+p_2^*{\cal O}_{\P^2}(3)$, so $D_1^2=0$, $D_2^2=2$ and $D_1\cdot D_2=3$. Therefore, the generic N\'eron-Severi lattice of hypersurfaces of bidegree (2,3) in $\P^1\times \P^2$ has intersection matrix
\begin{equation}
\begin{pmatrix}
0&3\\
3&2\\
\end{pmatrix}.
\end{equation}
The divisor class $D_1$ represents a fiber class, and the divisor class $D_2$ is a 3-section.  
\par When an elliptic surface has a section, let us denote a divisor class of section by $O$ and fiber class by $F$ respectively, then section $O$ and fiber class $F$ satisfy the relation $O\cdot F=1$. The divisor class $D_1$ represents a fiber class. (So it deserves another notation $F$.) But the N\'eron-Severi lattice of a generic hypersurface $S$ in $\P^1\times \P^2$ constructed above has rank 2, generated by divisors $D_1$ and $D_2$. So any divisor $D$ in $S$ can be written as an integral sum of $D_1$ and $D_2$, $D=nD_1+mD_2$. Then for any divisor $D$ in $S$, $D\cdot D_1(=D\cdot F)=3m\ne 1$. Therefore, generic member of a bidegree (2,3) hypersurface $S$ in $\P^1\times \P^2$ does {\it not} have a section. (The above argument holds true only for a {\it generic} choice of an equation for $S$; there are exceptions. Special cases such as a hypersurface $S$ is given by the equation $T_1^2ZY^2=T_1^2X^3+T_0^2XZ^2+T_0^2Z^3$, where $[X:Y:Z]$ is a coordinate on $\P^2$ and $[T_0:T_1]$ is a coordinate on $\P^1$, should be excluded. Such an equation locally becomes $y^2=x^3+t^2x+t^2$, where $x=X/Z, y=Y/Z$ and $t=T_0/T_1$. An equation of Weierstrass form admits a constant section $[0:1:0]$ over the base. Therefore, to construct a K3 surface without section, we need to avoid a hypersurface given by an equation which admits a transformation into Weierstrass form.) 
\par In subsequent sections, we perform a detailed analysis of F-theory on an elliptic Calabi-Yau 4-fold without section. We restrict our consideration to F-theory compactification on a product an elliptic K3 surface times a K3 surface, with an elliptic K3 surface chosen from hypersurfaces of bidegree (2,3) in $\P^1\times \P^2$ constructed above. As shown above, such chosen elliptic K3 surface does not have a section. Therefore, K3 $\times$ K3, with one of the K3's chosen from hypersurfaces of bidegree (2,3) in $\P^1\times \P^2$ constructed above, does not have section as an elliptic Calabi-Yau 4-fold. 
\par To study F-theory compactifications on a class of K3 $\times$ K3 without a section in detail, we limit ourselves to consider two specific families of K3 surfaces\footnote{For a generic hypersurface $S$ of bidegree (2,3) in $\P^1\times\P^2$, Picard number $\rho(S)=2$. For some specific surfaces, Picard number may enhance. This means divisors increase, but it is expected that divisors increase only in the fiber direction. Therefore, basic line of the above argument holds true for such specific K3 surfaces with enhanced Picard numbers, and they still do not have section.}, among hypersurfaces of bidegree (2,3) in $\P^1\times\P^2$. One family consists of K3 surfaces of {\it Fermat type} defined by equations of the form:
\begin{equation}
fX^3+gY^3+hZ^3=0.
\label{eq:first}
\end{equation}
$[X:Y:Z]$ is a coordinate on $\P^2$, and $f,g,h$ are homogeneous polynomials of degree 2 in $[T_0:T_1]$ where $[T_0:T_1]$ is a coordinate on $\P^1$. For notational simplicity we use $t:=T_0/T_1$. Then we may rewrite $f, g, h$ as polynomials in $t$ of degree at most 2. (Original homogeneous forms can be easily recovered.) 
\par Notice that an elliptically fibered hypersurface having a section over the base $\P^1$ means that an equation for a hypersurface has a $K(\P^1)-$rational point. Here $K(\P^1)$ denotes a function field over $\P^1$. (So $K(\P^1)\cong \C(t)$.) For generic polynomials $f,g,h$ of degree 2 in $t$, the K3 surface determined by (\ref{eq:first}) does not admit a section, i.e. the equation (\ref{eq:first}) with variables $X,Y,Z$ does not have a solution in $\C(t)$. 
\par We rewrite the equation (\ref{eq:first}) as 
\begin{equation}
(t-\alpha_1)(t-\alpha_2)X^3+(t-\alpha_3)(t-\alpha_4)Y^3+(t-\alpha_5)(t-\alpha_6)Z^3=0
\label{eq:choice}
\end{equation} 
where $\alpha_i\in\C$, $\alpha_1$ and $\alpha_2$ are the roots of $f$, $\alpha_3$ and $\alpha_4$ are the roots of $g$, and $\alpha_5$ and $\alpha_6$ are the roots of $h$\footnote{We suppressed top coefficients as they are irrelevant to locations and types of singular fibers, so are irrelevant to gauge groups.}. Then the K3 surface given by the equation (\ref{eq:choice}) has singular fibers at $t=\alpha_i$, $i=1,\cdots, 6$. A special case in which some $\alpha_i$, say $\alpha_1$ is $\infty$ corresponds to the equation
\begin{equation}
(t-\alpha_2)X^3+(t-\alpha_3)(t-\alpha_4)Y^3+(t-\alpha_5)(t-\alpha_6)Z^3=0.
\end{equation} This way, reduction of a factor in $t$ from the coefficients of $X,Y,Z$ in (\ref{eq:choice}) results in singular fiber at $\infty\in\P^1$. 
\par One needs to be careful about the choice of $\{\alpha_i\}^6_{i=1}$; a surface given by the equation (\ref{eq:choice}) develops singularity worse than a rational double point with a wrong choice of $\{\alpha_i\}^6_{i=1}$. This bad singularity ruins the triviality of canonical bundle, and the equation (\ref{eq:choice}) does not give a K3 surface then.  
\par A simple condition on the multiplicity of $\alpha_i$'s in $\{\alpha_i\}^6_{i=1}$ ensures that the equation (\ref{eq:choice}) correctly gives a K3 surface. The rule is that any two $\alpha_j$ and $\alpha_k$ among 6 $\alpha_i$'s may coincide, but any triplet among 6 $\alpha_i$'s are not allowed to coincide. In other words, the equation (\ref{eq:choice}) describes a K3 surface if and only if the multiplicity of $\alpha_i$ is at most 2 for every $i=1,\cdots,6$. For instance, the equation (\ref{eq:choice}) with $\alpha_1=\alpha_2=p$, $\alpha_3=\alpha_4=q$, $\alpha_5=\alpha_6=r$, where $p,q,r$ are mutually distinct, correctly describes a K3 surface, but the one with $\alpha_1=\alpha_2=\alpha_3$ does not give a K3 surface.    
\par In addition to the above K3 surfaces defined by Fermat type equation (\ref{eq:choice}), we also would like to consider another family of K3 surfaces defined by equation in {\it Hesse form}:
\begin{equation}
aX^3+bY^3+cZ^3-3dXYZ=0,
\label{eq:second}
\end{equation}
where $a,b,c,d$ are homogeneous polynomials in $[T_0:T_1]$ of degree 2. For generic polynomials $a,b,c,d$, the above equation of Hesse form (\ref{eq:second}) does not have a $K(\P^1)-$rational point, i.e. a K3 surface defined by the equation (\ref{eq:second}) does not have a section. Again for notational simplicity, we rewrite $a,b,c,d$ as polynomials in $t$ of degree at most 2. 
\par In section 3, we determine gauge groups arising on 7-branes for F-theory compactified on K3 $\times$ K3, with the one of K3's being a K3 surface of Fermat type defined by the equation (\ref{eq:choice}), and being a K3 surface in Hesse form defined by the equation (\ref{eq:second}), by computing their singular fibers. These K3 surfaces do not have a section, so the gauge groups which arise on 7-branes do not have $U(1)$-factor.     

\section{Gauge Groups on the Model}
In this section, we compute gauge groups which arise on 7-branes for F-theory compactification on an elliptic K3 surface times a K3 surface. For an elliptic K3 surface in the product K3 $\times$ K3, we choose a K3 surface defined by the equation of Fermat type
\begin{equation}
(t-\alpha_1)(t-\alpha_2)X^3+(t-\alpha_3)(t-\alpha_4)Y^3+(t-\alpha_5)(t-\alpha_6)Z^3=0,
\label{eq:choice 2}
\end{equation} 
and a K3 surface defined by the equation in Hesse form
\begin{equation}
aX^3+bY^3+cZ^3-3dXYZ=0,
\label{eq:second 2}
\end{equation}
as discussed in the previous section. With these settings, the product K3 $\times$ K3 does not have a section. So, the gauge groups arising on 7-branes for F-theory on these models do not have $U(1)$-factor. 
\par Fiber of an elliptic manifold is generically a smooth elliptic curve, but along a locus of codimension 1 in the base a fiber becomes singular. This locus is called the discriminant locus.  
\par The singular fibers of an elliptic surface were classified by Kodaira\cite{Kod}. We use Kodaira's notation for types of singular fibers. Singular fiber of an elliptic surface is either the sum of smooth $\P^1$'s intersecting in specific ways, or a $\P^1$ with one singularity. The latter case is not a singularity of a manifold itself. When a singular fiber is $\P^1$ with a single singularity, it is either a nodal rational curve denoted by $I_1$, or a cuspidal rational curve denoted by type $II$, respectively in Kodaira's notation. When singular fiber is reducible into the sum of smooth $\P^1$'s, there are seven types of singular fibers: two infinite series $I_n$ and $I^*_n$, and five types $III, IV, II^*, III^*$ and $IV^*$. 
\par
\begin{figure}
\begin{center}
\includegraphics[height=6cm, bb=0 0 1003 801]{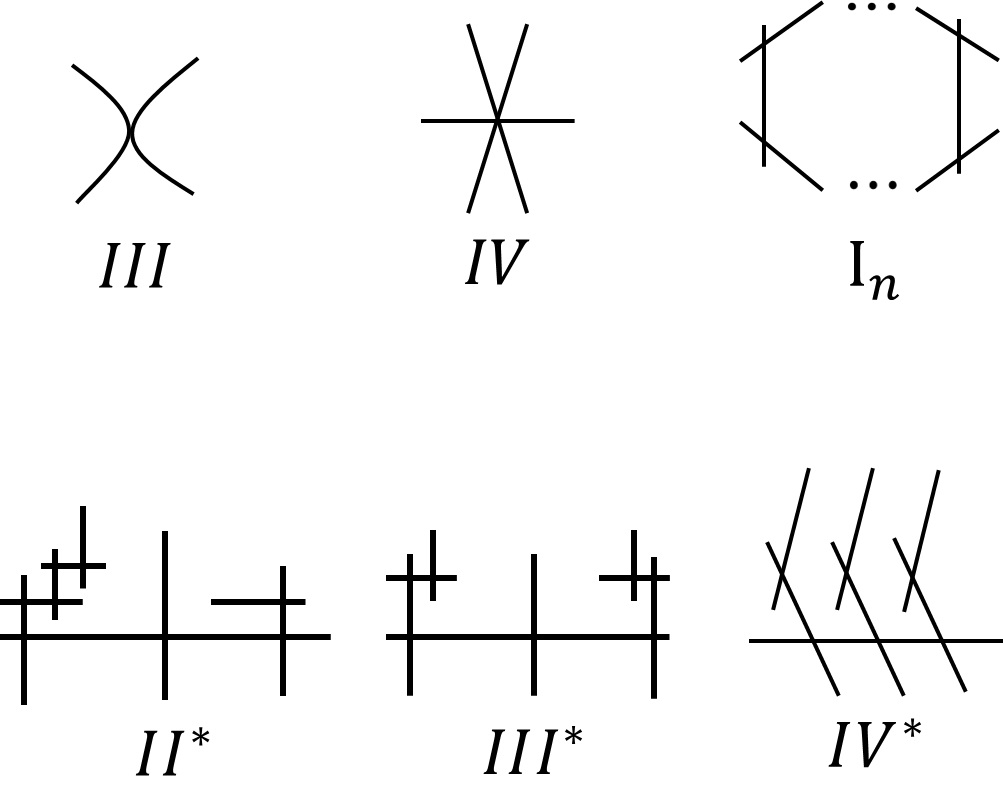}
\caption{\label{fig}Singular Fibers}
\end{center}
\end{figure}

 To each singular fiber an extended Dynkin diagram is associated with a vertex for each $\P^1$ component of the fiber. Two vertices are joined by an edge if and only if the corresponding $\P^1$'s intersect. Extended Dynkin diagrams $\til A_{n-1}$, $\til D_{n+4}$, $\til E_6$, $\til E_7$ and $\til E_8$ describe the configurations of $\P^1$'s of singular fibers of types $I_n$, $I^*_n$, $IV^*$, $III^*$ and $II^*$ respectively. Type $III$ is two $\P^1$'s tangent to each other at a point, and type $IV$ is three $\P^1$'s meeting at a single point. Extended Dynkin diagrams associated to fiber types $III$ and $IV$ are $\til A_1$ and $\til A_2$ respectively. For pictures of singular fibers, see Figure \ref{fig}. Each line in a picture represents a $\P^1$ component. Pictures show the configurations of $\P^1$ components of each fiber type and how they intersect one another.   
\par Extended Dynkin diagrams $\til A_n$, $\til D_n$ and $\til E_n$ correspond to $SU(n+1)$, $SO(2n)$ and $E_n$ gauge groups respectively, on 7-branes. So, singular fibers of types $I_n$, $I^*_n$ give rise to $SU(n)$ and $SO(2n+8)$ gauge groups. Types $IV^*$, $III^*$ and $II^*$ generate $E_6$, $E_7$ and $E_8$ gauge groups respectively. Singular fibers of types $I_1$ and $II$ do not generate non-Abelian gauge groups on 7-branes. Type $III$ fiber gives $SU(2)$ gauge group, and type $IV$ generates $SU(3)$ gauge group. The correspondence between the fiber types and the gauge groups are shown in the Table 1 below. 

\begingroup
\renewcommand{\arraystretch}{1.5}
\begin{table}[htb]
\begin{center}
  \begin{tabular}{|c|c|} \hline
fiber type & gauge group \\ \hline
$I_n$ & $SU(n)$ \\
$I^*_n$ & $SO(2n+8)$ \\
$III$ & $SU(2)$ \\
$IV$ & $SU(3)$ \\
$IV^*$ & $E_6$ \\
$III^*$ & $E_7$ \\
$II^*$ & $E_8$ \\ \hline   
\end{tabular}
\caption{The correspondence between the fiber types and the gauge groups.}
\end{center}
\end{table}  
\endgroup 
\par In the following subsections, we determine gauge groups on 7-branes for F-theory on a K3 surface of Fermat type times K3, and a K3 surface in Hesse form times K3, respectively. 
\par After we deduce gauge groups on 7-branes for F-theory compactifications on K3 $\times$ K3 without a section, we make a comment on possible monodromies. We find that for a Fermat type K3 $\times$ a K3, it is essential to consider monodromies to determine the gauge symmetries. 
\par We also discuss anomaly cancellation conditions on the models. F-theory compactification on K3 $\times$ K3 gives 4d theory with enhanced $N=2$ supersymmetry. Resultantly, theory is highly constrained by consistency conditions from anomalies. We check if the solutions we obtain are consistent with anomaly cancellation conditions. We also see that the form of the discriminant locus is determined by the consistency conditions from anomalies.     

\subsection{Gauge Groups on K3 surfaces of Fermat Type (times K3)}
In this subsection we determine gauge groups arising on 7-branes for F-theory compactification on K3 $\times$ K3, with one of the K3's given by the equation of Fermat type (\ref{eq:choice 2}). 
\par As we saw in the previous section, K3 surface of Fermat type does not admit a section. This means that the equation (\ref{eq:choice 2}) does not transform into Weierstrass form. So we cannot use Tate's algorithm\cite{Tate} to determine singular fibers directly from the defining equation. 
\par So we consider Jacobian fibration $J(S)$ of K3 surface of Fermat type $S$ to compute singular fibers. Jacobian $J(S)$ of $S$ is an elliptic surface constructed from $S$ with exactly the same types of singular fibers at exactly the same locations over the base $\P^1$ as $S$. Even when $S$ does not have a section, $J(S)$ has a section, so $J(S)$ does admit transformation into Weierstrass form. 
\par Jacobian $J(S)$ of K3 surface $S$ defined by (\ref{eq:choice 2}) is given by the equation 
\begin{equation}
X^3+Y^3+\Pi^6_{i=1}(t-\alpha_i)Z^3=0,
\label{eq:jacobian choice}
\end{equation} 
where $t$ runs through $\P^1$. Converting the equation (\ref{eq:jacobian choice}) by a linear change of variables to the Weierstrass form and applying Tate's algorithm, we find the locations of singular fibers over the base $\P^1$ and their types of Jacobian $J(S)$. Both Jacobian $J(S)$ and the original surface $S$ have the same positions and types of singular fibers. So the computational results for the Jacobian $J(S)$ precisely tell us the locations and types of singular fibers of the original K3 surface $S$. 
\par The Weierstrass form into which the equation (\ref{eq:jacobian choice}) is transformed is locally given by the equation
\begin{equation}
y^2=x^3-2^4\cdot 3^3\cdot \Pi^6_{i=1}(t-\alpha_i)^2. 
\end{equation}
Therefore the discriminant of the Jacobian $J(S)$ is given by
\begin{equation}
\Delta=\Pi^6_{i=1}(t-\alpha_i)^4. 
\label{eq:Fermat discriminant}
\end{equation}
\par We can determine the locations and types of the singular fibers of the Jacobian $J(S)$ from the discriminant $\Delta$ given by (\ref{eq:Fermat discriminant}). Singular fibers are at $t=\alpha_i$, $i=1,\cdots, 6$ and they are seen to be additive. Using Tate's algorithm, we see that when $\alpha_i\ne\alpha_j$ for all $j\ne i$, the singular fiber has type $IV$ at $t=\alpha_i$. When there is a single $j\ne i$ such that $\alpha_i=\alpha_j$, the singular fiber at $t=\alpha_i$ has type $IV^*$. Intuitively speaking, when two points $\alpha_i$ and $\alpha_j$ on $\P^1$ approach each other and they eventually coincide (i.e. $\alpha_i=\alpha_j$), two singular fibers of type $IV$ at $t=\alpha_i$ and $t=\alpha_j$ merge to give a singular fiber of type $IV^*$. As noted in section 2, more than two $\alpha_i$'s cannot be coincident. When more than two $\alpha_i$'s coincide (for instance $\alpha_i=\alpha_j=\alpha_k$ with $i,j,k$ mutually distinct), the equation (\ref{eq:choice 2}) does not give a K3 surface.  
\par In this way we find that singular fibers of Fermat type K3 surface $S$ defined by the equation (\ref{eq:choice 2}) are located at $t=\alpha_i$, $i=1,\cdots,6$, and when $\alpha_i$'s are mutually distinct, their fiber types are all $IV$. So K3 surface $S$ of Fermat type generically has 6 singular fibers of type $IV$. When two of $\alpha_i$'s coincide, two fibers of type $IV$ collide and enhance to a singular fiber of type $IV^*$. Therefore, we have 4 cases of the configurations of the singular fibers in total: (i)six fibers of type $IV$, (ii)four fibers of type $IV$ and one fiber of type $IV^*$, (iii)two fibers of type $IV$ and two fibers of type $IV^*$, and (iv)three fibers of type $IV^*$. 
\par Summarising the above, we have 4 cases of gauge groups which arise on 7-branes. They are:
\begin{itemize}
\item $SU(3)^{\oplus6}$
\item $E_6\times SU(3)^{\oplus4}$
\item $E_6^{\oplus2}\times SU(3)^{\oplus2}$
\item $E_6^{\oplus3}$.
\end{itemize}
The gauge group which arises on 7-branes is generically $SU(3)^{\oplus6}$ for F-theory on K3 $\times$ K3, with one of the K3's being a K3 surface of Fermat type $S$. But when two of $\alpha_i$'s coincide, two $SU(3)$'s collide and enhance to the exceptional gauge group $E_6$, and we have more $E_6$'s as we have more pairs of coincident $\alpha_i$'s. As mentioned in the previous section, when three of $\alpha_i$'s coincide, the singularity at the point of triple coincidence becomes worse than a rational double point. This is not allowed for the equation (\ref{eq:choice 2}) to describe a K3 surface.    
\par The above 4 cases classify the gauge groups which arise on 7-branes for F-theory on an elliptic K3 surface of Fermat type times a K3 surface. The product K3 $\times$ K3, with one of the K3's being Fermat type, does not have a section as an elliptic Calabi-Yau 4-fold. Therefore the gauge groups on 7-branes do not have $U(1)$-factor.

\subsection{Gauge Groups on K3 surfaces of Hesse Form (times K3)}
In this subsection, we investigate gauge groups arising on 7-branes when F-theory is compactified on K3 $\times$ K3, with one of the K3's being a K3 surface of Hesse form defined by the equation (\ref{eq:second 2}). K3 surface $S$ of Hesse form does not have a section for generic polynomials $a,b,c$ and $d$ with degrees 2 in $t$. 
\par By inspection on the condition that the equation (\ref{eq:second 2}) has a singularity, we can directly compute the discriminant of the equation (\ref{eq:second 2}). We denote by $\Delta$ the discriminant of the equation (\ref{eq:second 2}). For notational convenience, we set 
\begin{equation}
f_{Hesse}:=aX^3+bY^3+cZ^3-3dXYZ.
\end{equation}
Then the equation (\ref{eq:second 2}) has a singularity precisely when the equations
\begin{equation}
\partial_X f_{Hesse}=\partial_Y f_{Hesse}=\partial_Z f_{Hesse}=0
\label{eq:Hesse singularity}
\end{equation}
have a solution. From the equations in (\ref{eq:Hesse singularity}), we deduce that the discriminant $\Delta$ contains the factors $abc$ and $(abc-d^3)$. Then by considering some special case of the equation (\ref{eq:second 2}), we see that the multiplicity of the factor $(abc-d^3)$ in the discriminant $\Delta$ is 3. Both $\Delta$ and $abc(abc-d^3)^3$ have degree 24 as polynomials in t, so we can conclude that they are equal. Therefore, the discriminant $\Delta$ is given by   
\begin{equation}
\label{eq:Delta}
\Delta= abc(abc-d^3)^3.
\end{equation}
(We suppressed an irrelevant constant factor of the discriminant.)
\par $abc-d^3$ is a polynomial in $t$ of degree 6, so we can rewrite the equation (\ref{eq:Delta}) as  
\begin{equation}
\Delta=abc\cdot \Pi^6_{i=1}(t-\gamma_i)^3,
\label{eq:disc}
\end{equation}
where $\gamma_i$'s are six roots of $abc-d^3$. From the expression of the discriminant (\ref{eq:disc}), we see that singular fibers are at $t=\gamma_i$ $(i=1,\cdots,6)$, and at zeros of $a,b$ and $c$. 
\par It can be seen that every fiber is multiplicative. So each singular fiber has fiber type $I_n$ for some $n$, where $n$ equals the number of irreducible components of a singular fiber. When there is a multiplicative singular fiber at $t=t_0$, the number of the components of the singular fiber at $t=t_0$ is equal to the multiplicity of the discriminant $\Delta$ at $t=t_0$. In this way, we can determine the types of singular fibers from the expression for the discriminant $\Delta$ (\ref{eq:disc}). 
\par When $\gamma_i$'s are mutually distinct, the multiplicity of each $\gamma_i$ in the discriminant $\Delta$ is 3, so all the six fibers at $t=\gamma_i$ $(i=1,\cdots,6)$ are of type $I_3$. When two of $\gamma_i$'s coincide, the multiplicities of the two add up to 6 so the fiber has type $I_6$ at the coincident $\gamma_i$. Intuitively speaking, two fibers of type $I_3$ collide and enhance to type $I_6$ fiber. When $a$ does not have a multiple root, say $a=const.(t-\beta_1)(t-\beta_2)$, $\beta_1\ne\beta_2$, then singular fibers at $t=\beta_i$, $i=1,2$ both have type $I_1$. These type $I_1$ fibers do not correspond to a surface singularity, and they do not generate non-Abelian gauge groups on 7-branes. When $a$ has a multiple root, i.e. when $\beta_1=\beta_2$, two singular fibers of type $I_1$ collide and enhance to a singular fiber of type $I_2$. The same argument applies to $b$ and $c$. When the polynomial $abc$ has a multiple root of multiplicity 3, say $abc$ has a triple root $t=\beta$ so $(t-\beta)^3$ divides $abc$, then there is a singular fiber of type $I_3$ at $t=\beta$.
\par Summarising the above, the gauge group on 7-branes is generically $SU(3)^{\oplus6}$ for F-theory on K3 $\times$ K3, with one of the K3's being a K3 surface of Hesse form. When $abc$ has a multiple root, the gauge group enhances to $SU(3)^{\oplus6}\times SU(2)$ when the multiplicity of the root is 2, and it enhances to $SU(3)^{\oplus7}$ when the multiplicity is 3. When two of $\gamma_i$'s coincide in the discriminant $\Delta$ given by the equation (\ref{eq:disc}), two $SU(3)$'s collide and enhance to the group $SU(6)$. From the expression for the discriminant $\Delta$ (\ref{eq:disc}), all the singular fibers are seen to be multiplicative. So, for a K3 surface of Hesse form, each singular fiber has type $I_n$ for some $n$. Therefore, only $SU(N)$ gauge groups arise on 7-branes for this model. $SO(2N)$ gauge groups and exceptional gauge groups do not arise on 7-branes for F-theory on a K3 surface of Hesse form times a K3 surface. This determines the gauge groups arising on 7-branes for F-theory compactified on an elliptic K3 surface in Hesse form given by the equation (\ref{eq:second 2}) times a K3 surface. The product K3 $\times$ K3, with one of the K3's being a K3 surface in Hesse form, does not have a section as an elliptic Calabi-Yau 4-fold. Therefore the gauge groups on 7-branes do not have $U(1)$-part. 

\subsection{Consideration on Monodromy and Consistency Conditions from Anomalies}
In this subsection, we comment on the monodromies in the gauge groups arising on 7-branes. We find that for F-theory on a Fermat type K3 surface times a K3 surface, considering monodromy uniquely determines the gauge groups which arise on 7-branes. We also check if the results we obtained above meet the consistency conditions from anomaly cancellation. 
\par We first see that consideration on the j-invariant greatly constrains the possible types of singular fibers for a Fermat type K3 surface. Every smooth fiber of a Fermat type K3 surface over the base is the Fermat curve. The j-invariant of the Fermat curve is known to be 0. Therefore, a fiber of a Fermat type K3 surface has j-invariant 0 throughout the base. This forces j-invariants of singular fibers of a Fermat type K3 surface to be 0. 
\par The j-invariant of each fiber type was computed by Kodaira in \cite{Kod}. We summarise his results in Table 2. ``Finite'' for the j-invariant of fiber type $I^*_0$ in the table means that the j-invariant takes a finite value for fiber type $I^*_0$ (against the infinity). The value of j-invariant of fiber type $I^*_0$ depends on the situations. So fiber type $I^*_0$ may have the j-invariant 0. Only singular fibers of type $II$, $II^*$, $IV$ and $IV^*$ have j-invariant 0. Fiber type $I^*_0$ can have j-invariant 0. Therefore, we deduce that these 5 are the only possible types of singular fibers for a Fermat type K3 surface. Equivalently, possible gauge groups on 7-branes for F-theory on a Fermat type K3 surface times a K3 surface are: $SU(3)$, $SO(8)$, $E_6$ and $E_8$. (Type $II$ fiber does not generate a non-Abelian gauge group on 7-branes.) We saw the above that the gauge groups which arise on 7-branes for this model are $SU(3)$ and $E_6$. So we confirm that our results obtained above agree with the requirement from a consideration on j-invariants. 
\par Since j-invariant of elliptic curve given by the equation of Hesse form varies over the base $\P^1$, we cannot apply the above argument to F-theory on a K3 surface of Hesse form times a K3 surface. 
\par $SL_2(\Z)$ has non-trivial finite subgroups of orders only 2, 3, 4 and 6. They are all cyclic: $\Z_2$, $\Z_3$, $\Z_4$ and $\Z_6$. It follows that there are only finite order elements of $SL_2(\Z)$ with orders 1,2,3,4 and 6. Singular fibers of fiber type $I^*_0$, $II$, $III$, $IV$, $II^*$, $III^*$ and $IV^*$ have finite order monodromies, represented by these elements. For monodromies of fiber types and their orders, see Table 2. Results which appear in Table 2 were derived by Kodaira\cite{Kod}\footnote{Kodaira\cite{Kod} computed Euler numbers of singular fibers. Euler number of a singular fiber can be interpreted as the number of 7-branes associated to the fiber type of a singular fiber.}. Singular fibers of type $II$ and $II^*$ have monodromies of finite order 6, while the fibers of type $IV$ and $IV^*$ have monodromies of order 3. Therefore, from a consideration on j-invariant we learn that possible monodromies for singular fibers of a Fermat type K3 surface have order 3 or 6. Combined with consistency conditions from anomalies, below we will see that only the monodromies of order 3 occur for singular fibers of a Fermat type K3 surface. 
\begingroup
\renewcommand{\arraystretch}{1.1}
\begin{table}[htb]
  \begin{tabular}{|c|c|r|c|c|} \hline
fiber type & j-invariant & Monodromy  & order of Monodromy & \# of 7-branes (Euler number) \\ \hline
$I^*_0$ & finite & $-\begin{pmatrix}
1 & 0 \\
0 & 1 \\
\end{pmatrix}$ & 2 & 6\\ \hline
$I_b$ & $\infty$ & $\begin{pmatrix}
1 & b \\
0 & 1 \\
\end{pmatrix}$ & infinite & $b$\\
$I^*_b$ & $\infty$ & $-\begin{pmatrix}
1 & b \\
0 & 1 \\
\end{pmatrix}$ & infinite & $b+$6\\ \hline
$II$ & 0 & $\begin{pmatrix}
1 & 1 \\
-1 & 0 \\
\end{pmatrix}$ & 6 & 2\\
$II^*$ & 0 & $\begin{pmatrix}
0 & -1 \\
1 & 1 \\
\end{pmatrix}$ & 6 & 10\\ \hline
$III$ & 1728 & $\begin{pmatrix}
0 & 1 \\
-1 & 0 \\
\end{pmatrix}$ & 4 & 3\\
$III^*$ & 1728 & $\begin{pmatrix}
0 & -1 \\
1 & 0 \\
\end{pmatrix}$ & 4 & 9\\ \hline
$IV$ & 0 & $\begin{pmatrix}
0 & 1 \\
-1 & -1 \\
\end{pmatrix}$ & 3 & 4\\
$IV^*$ & 0 & $\begin{pmatrix}
-1 & -1 \\
1 & 0 \\
\end{pmatrix}$ & 3 & 8\\ \hline
\end{tabular}
\caption{Fiber types and their j-invariants, monodromies, and associated numbers of 7-branes.}
\end{table}
\endgroup
\par Now we consider the consistency conditions from anomalies on our models. We will see that the above consideration on j-invariants together with consistency conditions from anomalies precisely determines the gauge groups on 7-branes for F-theory on a Fermat type K3 surface times a K3 surface. 
\par F-theory compactification on K3 $\times$ K3 gives 4d theory with $N=2$ supersymmetry. The formula for the obstruction to cancelling the tadpole anomaly by turning on branes, for type IIA, M-theory and F-theory on a Calabi-Yau 4-fold, was derived in \cite{SVW}. This anomaly cancellation is required to preserve supersymmetry of the theory. In particular, for F-theory on K3 $\times$ K3, the tadpole can be cancelled by turning on $N$ 3-branes, where
\begin{equation} 
N=\chi(K3 \times K3)/24=24.
\end{equation}
(They also derived some topological constraints on the formula for an elliptic Calabi-Yau 4-fold, with assumptions that an elliptic Calabi-Yau 4-fold has a section and admits a smooth Weierstrass model. So the topological constraints they derived do not apply to our model.) 
\par When 24 3-branes are turned on, F-theory has 3-brane charge 24. For a product K3 $\times$ K3 (with one of the K3's being elliptically fibered), the base is $\P^1\times$ K3. Codimension-1 locus of $\P^1$ is set of discrete points, so each 7-brane is wrapped on $\{{\rm pt}\}\times$ K3, i.e. a 7-brane is wrapped on a K3 surface. Each 7-brane wrapped on a K3 surface produces 3-brane charge $-1$. Since 3-brane charge from 7-branes wrapped on K3 surfaces cancel the charges from 3-branes and the net 3-brane charge is 0, we learn that there are 24 7-branes in total. Therefore, we see that the tadpole anomaly cancellation condition determines the form of the discriminant locus to be \{24 points (counted with multiplicity)\} $\times$ K3. Here the counting of points is done with weight of multiplicity assigned, so the actual number of points can be less than 24. 
\par Now we combine the requirement from the consideration on j-invariants above with the consistency condition from anomalies to deduce gauge groups on 7-branes for F-theory on a Fermat type K3 surface times a K3 surface. We find that with these constraints, the gauge groups are precisely determined. 
\par From the form of the equation (\ref{eq:choice 2}), we see that a fiber degenerates precisely when $t=\alpha_i$, $i=1,\cdots,6$. So we expect that the six points $\{\alpha_i\}^6_{i=1}$ on $\P^1$ give all the locations of singular fibers. When $\alpha_i$'s are mutually distinct, there are 6 singular fibers. We have 24 7-branes in total, and by a symmetry argument there should be 4 7-branes at each singular fiber. From the consideration on j-invariant we learned that the possible types of singular fibers are $I^*_0$, $II$, $II^*$, $IV$ and $IV^*$. From Table 2, we see that only the fiber type $IV$ corresponds to 4 7-branes. Therefore we can conclude that when $\alpha_i$'s are mutually distinct, singular fiber at each $t=\alpha_i$ has type $IV$, in agreement with what we obtained above. 
\par To determine the fiber type at $t=\alpha_j$ when $\alpha_j$ and $\alpha_k$ coincide for some $k\ne j$, we consider the special case of the equation (\ref{eq:choice 2}):
\begin{equation}
(t-\alpha_1)^2X^3+(t-\alpha_3)^2Y^3+(t-\alpha_5)^2Z^3=0,
\label{eq:special fermat type}
\end{equation}
so we can apply a symmetry argument. From the form of the equation (\ref{eq:special fermat type}), we see that singular fibers are $t=\alpha_i$, $i=1,3,5$, and they have the same fiber type. So there are 3 singular fibers of the same type, and there should be 8 7-branes at each fiber. As mentioned in the previous paragraph, since singular fibers have j-invariant 0, possible types of a singular fiber are type $I^*_0$, $II$, $II^*$, $IV$ and $IV^*$ for a Fermat type K3 surface. From Table 2, we see that the only fiber type $IV^*$ corresponds to 8 7-branes. Therefore we learn that when two of $\alpha_i$'s are coincident, say $\alpha_j=\alpha_k$, then singular fiber at $t=\alpha_j$ has type $IV^*$. So we recover the results that we obtained the above.
\par For F-theory compactification on a K3 in Hesse form $\times$ a K3, we see that the expression (\ref{eq:Delta}) for the discriminant $\Delta$ of a K3 in Hesse form has degree 24 in the variable $t$. This corresponds to the physical condition that there are in total 24 7-branes wrapped on the components of the discriminant locus. Therefore, we can conclude that the solutions for the gauge groups on 7-branes we obtained the above are in accord with the anomaly cancellation conditions for F-theory compactification on a K3 in Hesse form $\times$ a K3 as well.
\par For F-theory on K3 $\times$ K3, the above consistency conditions from anomaly cancellation corresponds to the mathematical fact that the sum of Euler numbers of singular fibers equals 24 for an elliptic K3 surface. Euler number of a singular fiber is equal to the number of 7-branes needed to generate a gauge group associated to the fiber type. This mathematical fact about elliptic K3 holds true, regardless of whether an elliptic K3 surface admits a section or not. 
\par Fiber types $IV$ and $IV^*$ are the unique fiber types with monodromies of order 3. Fiber types $II$ and $II^*$ are the unique fiber types with monodromies of order 6. Consideration on j-invariants limited possible fiber types to the ones with monodromies of finite orders 3 and 6. The consideration on j-invariant, combined with tadpole anomaly cancellation condition precisely determined the types of singular fiber of a Fermat type K3 surface to be type $IV$ and $IV^*$. Type $IV$ and $IV^*$ are the exact fiber types with monodromies of order 3. So the monodromies of order 3 characterise singular fibers for F-theory on a Fermat type K3 surface times a K3 surface. Therefore we see that for a Fermat type K3 surface, it is essential to consider a monodromy group in determining the gauge symmetries.

\section{Matter Fields on 7-branes}
In this section, we investigate matter fields on 7-branes for F-theory compactified on K3 $\times$ K3, with one of the K3's chosen to be a Fermat type K3 surface or a K3 surface in Hesse form. With these choices, K3 $\times$ K3 does not have a section. 
\par For F-theory on K3 $\times$ K3, all the 7-branes are parallel, so we do not have Yukawa-type interactions for our models. F-theory compactification on K3 $\times$ K3 gives a 4d theory with $N=2$ supersymmetry in absence of flux. By including flux, half the supersymmetry is broken, and theory becomes 4d with $N=1$ supersymmetry. Since all the 7-branes are parallel, without fluxes the only light matter for these models are the adjoint matters of gauge groups on 7-branes. When fluxes are turned on, vector-like pairs could also arise. Including flux, however, potential matter may vanish due to anomalies. 
\par It turns out that, a Fermat type K3 surface with 3 singular fibers of type $IV^*$ is an attractive\footnote{It is standard to call a K3 surface with the highest possible Picard number $\rho=20$ a singular K3 surface in mathematics. We follow the convention of the term used in \cite{M}.} K3 (i.e. a K3 surface with highest possible Picard number $\rho=20$), and whose complex structure can be precisely determined. Therefore, some detailed analysis on the matter spectrum with flux is possible for F-theory on K3 $\times$ K3, with one of the K3's being a Fermat type with $E_6\times E_6\times E_6$ gauge group on 7-branes. For other cases, we can only say that vector-like pairs {\it could} arise, but they may in fact vanish. The net chirality of light matter arising on 7-branes for F-theory compactification on K3 $\times$ K3 is 0\cite{BKW}. 
\par Now we see that for F-theory compactification on a Fermat type attractive K3 with gauge groups $E_6\times E_6\times E_6$ times a K3 (with K3 being appropriately chosen), tadpole anomaly cancels by turning on appropriate number of 3-branes. Consequently, we can calculate the full matter spectrum with flux for these cases.
\par Transcendental lattice $T_S$ of a K3 surface $S$ is the orthogonal complement of N\'eron-Severi lattice $NS$ in K3 lattice $H^2(S,\Z)$:
\begin{equation}
T_S:=NS^{\perp}\subset H^2(S,\Z).
\end{equation}
When a K3 surface $S$ is attractive (i.e. when the Picard number $\rho(S)=20$), its transcendental lattice $T_S$ is a positive definite even lattice of rank 2. \par Attractive K3 surfaces are classified by transcendental lattices, in the following sense. There is a bijective correspondence between complex structures of attractive K3 surfaces and the set of 2 $\times$ 2 even positive definite integral matrices modded out by the conjugacy action of $SL_2(\Z)$\cite{SI}. Let ${\cal Q}$ denote the set of 2 $\times$ 2 even positive definite integral matrices. Then $SL_2(\Z)$ acts on ${\cal Q}$ by 
\begin{equation}
Q \rightarrow g^t\cdot Q\cdot g,
\end{equation}
where $Q$ is in ${\cal Q}$ and $g$ is in $SL_2(\Z)$. So $Q$ and $Q'$ are identified in ${\cal Q}/SL_2(\Z)$ exactly when there is some $g$ in $SL_2(\Z)$ such that 
\begin{equation}
Q'=g^t\cdot Q\cdot g.
\end{equation} 
Each attractive K3 surface $S$ has the transcendental lattice $T_S$. Then with a basis $\{v_1, \, v_2\}$ of $T_S$, the transcendental lattice $T_S$ has an associated Gram matrix $Q_S$, 
\begin{equation}
Q_S:=\begin{pmatrix}
v_1\cdot v_1 & v_1\cdot v_2 \\
v_2\cdot v_1 & v_2\cdot v_2 \\
\end{pmatrix}=\begin{pmatrix}
2a & b \\
b & 2c \\
\end{pmatrix}
\end{equation}
for some integers $a$,$b$,$c$ in $\Z$. Then we can define the map $S\rightarrow Q_S$, assigning the Gram matrix $Q_S$ of the transcendental lattice to each attractive K3 $S$. Theorem 4 in \cite{SI} states that this map gives bijective correspondence between complex structures of attractive K3 surfaces and ${\cal Q}/SL_2(\Z)$. ($SL_2(\Z)$-action on ${\cal Q}$ corresponds to the change of basis of $T_S$.) In other words, the triplet of integers [$2a$ \hspace{1mm} $b$ \hspace{1mm} $2c$] (modded out by some $SL_2(\Z)$-action) parameterises the complex structure moduli of attractive K3 surfaces, and each fixed [$2a$ \hspace{1mm} $b$ \hspace{1mm} $2c$] can be identified with the transcendental lattice $T_S$ of an attractive K3 $S$ via
\begin{equation}
T_S=\begin{pmatrix}
2a & b \\
b & 2c \\
\end{pmatrix}.
\end{equation} 
\par Since a triplet [$2a$ \hspace{1mm} $b$ \hspace{1mm} $2c$] parameterises the complex structure moduli of attractive K3 surfaces, we use the symbol $S_{{\rm [}2a \hspace{1mm} b  \hspace{1mm} 2c{\rm]}}$ to represent an attractive K3 surface $S$, whose transcendental lattice $T_S$ has the Gram matrix $\begin{pmatrix}
2a & b \\
b & 2c \\
\end{pmatrix}$.
\par The transcendental lattice of an attractive Fermat type K3 surface with $E_6\times E_6\times E_6$ gauge groups (equivalently a Fermat type K3 surface with 3 singular fibers of type $IV^*$) has the Gram matrix $\begin{pmatrix}
6 & 3 \\
3 & 6 \\
\end{pmatrix}$
\footnote{We would like to thank Shigeru Mukai for pointing this out to us.}. So, $S_{[6 \hspace{1mm} 3 \hspace{1mm} 6]}$ represents an attractive Fermat type K3 surface with $E_6\times E_6\times E_6$ gauge groups, with the above notational convention. From lattice theoretic argument, N\'eron-Severi lattice $NS(S_{[6 \hspace{1mm} 3 \hspace{1mm} 6]})$ of the attractive Fermat type K3 surface $S_{[6 \hspace{1mm} 3 \hspace{1mm} 6]}$ is determined to be
\begin{equation}
NS(S_{[6 \hspace{1mm} 3 \hspace{1mm} 6]})=\begin{pmatrix}
0 & 3 \\
3 & 2 \\
\end{pmatrix} \oplus W,
\label{eq:ns lattice}
\end{equation}
where $W$ represents a (negative definite) overlattice of $E_6\oplus E_6\oplus E_6$ with discriminant 3 \footnote{We define $ADE$-lattice with negative sign in this paper.}. From the expression of N\'eron-Severi lattice (\ref{eq:ns lattice}), we can explicitly see that additional divisors $W$ increase only in the fiber direction (i.e. they are orthogonal to the lattice $\begin{pmatrix}
0 & 3 \\
3 & 2 \\
\end{pmatrix}$, which is spanned by fiber class and 3-section), and the attractive Fermat type K3 surface $S_{[6 \hspace{1mm} 3 \hspace{1mm} 6]}$ with fibration specified by gauge groups $E_6\times E_6\times E_6$ does not have a section. 
\par We will see below that, the full matter spectrum with flux can be calculated for F-theory compactification on $S_{[6 \hspace{1mm} 3 \hspace{1mm} 6]}$ $\times$ K3, with some appropriate choices of K3 surfaces. 
\par Flux compactification of M-theory on a product of attractive K3's $S_1$ $\times$ $S_2$ was considered in \cite{AK}. 4-form flux $G$ is subject to a quantisation condition\cite{W}
\begin{equation}
G \in H^4(S_1  \times  S_2, \Z),
\end{equation}
and has decomposition 
\begin{equation}
G=G_0+G_1,
\end{equation}
where
\begin{eqnarray}
G_0 & \in & H^{1,1}(S_1,\R)\otimes H^{1,1}(S_2,\R) \\
G_1 & \in & H^{2,0}(S_1,\C) \otimes H^{0,2}(S_2,\C)+{\rm h.c.}
\end{eqnarray}
\par In the presence of 4-form flux $G$, the tadpole cancellation condition \cite{SVW} becomes
\begin{equation}
\frac{1}{2}\int_{{\rm K3}\times {\rm K3}} G\wedge G+N_3=\frac{1}{24}\chi({\rm K3}\times {\rm K3})=24,
\label{eq:tadpole 4-form}
\end{equation}
where $N_3$ is the number of 3-branes turned on. With assumptions that 
\begin{equation}
G_0=0
\end{equation}
and
\begin{equation}
N_3=0,
\label{eq:flux assump}
\end{equation} 
\cite{AK} obtained all the pairs $S_{[2a \hspace{1mm} b \hspace{1mm} 2c]}$ $\times$ $S_{[2d \hspace{1mm} e \hspace{1mm} 2f]}$ which satisfy the tadpole cancellation condition (\ref{eq:tadpole 4-form}). 
Relaxing the condition (\ref{eq:flux assump}) to 
\begin{equation}
N_3\ge 0,
\end{equation}
the list of the pairs of attractive K3's was extended in \cite{BKW}. 
\par The lists in \cite{AK} and \cite{BKW} both contain only finitely many pairs of attractive K3's. So the complex structure moduli are the sets of finitely many discrete points. Therefore, the complex structure moduli in \cite{AK,BKW} are stabilised. The attractive K3 surface $S_{[6 \hspace{1mm} 3 \hspace{1mm} 6]}$ does not appear in the list of \cite{AK}, but it appears in the extended list of \cite{BKW}\footnote{\cite{BKW} uses different notation for attractive K3's. They denote by [a b c] the subscript for the attractive K3 surface whose transcendental lattice has the Gram matrix$\begin{pmatrix}
2a & b \\
b & 2c \\
\end{pmatrix}$. [3 3 3] in the list of \cite{BKW} represents the attractive K3 surface denoted by $S_{[6 \hspace{1mm} 3 \hspace{1mm} 6]}$ in this paper.}. 
\par 2 pairs $S_{[6 \hspace{1mm} 3 \hspace{1mm} 6]}$ $\times$ $S_{[4 \hspace{1mm} 2 \hspace{1mm} 4]}$ and $S_{[6 \hspace{1mm} 3 \hspace{1mm} 6]}$ $\times$ $S_{[2 \hspace{1mm} 1 \hspace{1mm} 2]}$ satisfy the tadpole cancellation condition (\ref{eq:tadpole 4-form}), with $N_3=6$ and $15$ respectively. So for F-theory compactifications on $S_{[6 \hspace{1mm} 3 \hspace{1mm} 6]}$ $\times$ $S_{[4 \hspace{1mm} 2 \hspace{1mm} 4]}$ and on $S_{[6 \hspace{1mm} 3 \hspace{1mm} 6]}$ $\times$ $S_{[2 \hspace{1mm} 1 \hspace{1mm} 2]}$, the tadpole anomalies are cancelled with appropriated number of 3-branes turned on. Therefore for these cases, we can say that the vector-like pairs {\it will} arise. 
Fermat type K3's with gauge groups $E_6\times E_6\times SU(3)^2$, $E_6\times SU(3)^4$ and $SU(3)^6$ are not attractive. For F-theory compactifications on these non-attractive K3 $\times$ a K3 or a K3 surface of Hesse form $\times$ a K3, we can only say that vector-like pair could arise with flux. 
\par As discussed in Katz and Vafa\cite{KV}, singularity of a space (on which theory is compactified) encodes information on charged matter. As 7-branes move apart from one another, they generate matter fields on 7-branes. In geometrical language, deformation of singularity of ADE-type (ADE-type of a singularity corresponds to  ADE-type of a singular fiber) describes this generation of matter. In subsections below, we describe light matter fields arising on 7-branes for F-theory compactified on a Fermat type K3 $\times$ a K3, and on a K3 surface of Hesse form $\times$ a K3, respectively. 
\subsection{Matter Fields for Fermat Type K3 surfaces (times K3)}
Recall we saw in section 3 that singular fibers of a Fermat type K3 surface have only types $IV$ or $IV^*$. (We classified all the 4 cases of configuration of singular fibers and gauge groups for F-theory compactifications on a Fermat type K3 $\times$ a K3.) Therefore, the corresponding surface singularities that a Fermat type K3 surface develops have types $A_2$ and $E_6$. 
\par F-theory compactification on K3 $\times$ K3 gives 4d theory with $N=2$ supersymmetry. All the 7-branes are parallel for F-theory on K3 $\times$ K3, so in absence of flux the only light matter are the adjoints of gauge groups on 7-branes. When fluxes are turned on, half the supersymmetry is broken, left with $N=1$ supersymmetry. Vector-like pairs from hypermultiplets could also arise with flux, but the vector-like pairs may vanish due to anomalies. As argued above, for F-theory compactifications on $S_{[6 \hspace{1mm} 3 \hspace{1mm} 6]}$ $\times$ $S_{[4 \hspace{1mm} 2 \hspace{1mm} 4]}$ and on $S_{[6 \hspace{1mm} 3 \hspace{1mm} 6]}$ $\times$ $S_{[2 \hspace{1mm} 1 \hspace{1mm} 2]}$, turning on appropriated number of 3-branes cancel the tadpole anomalies. For these models, vector-like pairs will arise by including flux.   
\par We compute the matter fields which can arise from $E_6$-singularity first. The only enhancement which generates the adjoint matters and hypermultiplets (an $N=2$ hypermultiplet is split into an $N=1$ vector-like pair with flux) is 
\begin{equation}
A_5\subset E_6.
\end{equation}
Under this enhancement, the adjoint of $E_6$ decomposes into irreducible representations of $A_5$ as
\begin{equation}
{\bf 78}={\bf 35}+{\bf 20}+\overline{\bf 20}+3\times{\bf 1},
\end{equation}
where {\bf 35} is the adjoint of $SU(6)$. The pair ${\bf 20}+\overline{\bf 20}$ forms an $N=2$ hypermultiplet. By including flux, the $N=2$ hypermultiplets are split into $N=1$ vector-like pairs ${\bf 20}+\overline{\bf 20}$. This vector-like pair ${\bf 20}+\overline{\bf 20}$ is only potential candidate for matter spectrum with flux, and the vector-like pairs may actually vanish due to anomalies.      
\par We next consider matters arising from $A_2$-singularity. For this case, the enhancement is 
\begin{equation}
A_1\subset A_2,
\end{equation}
and the adjoint of $A_2$ decomposes under this enhancement as  
\begin{equation}
{\bf 8}={\bf 3}+{\bf 2}+\overline{\bf 2}+{\bf 1},
\end{equation}
where {\bf 8} and {\bf 3} are the adjoints of $SU(3)$ and $SU(2)$, respectively. The pair ${\bf 2}+\overline{\bf 2}$ forms an $N=2$ hypermultiplet. By including flux, the $N=2$ hypermultiplets are split into the vector-like pairs ${\bf 2}+\overline{\bf 2}$. Again in general, we can only say that the vector-like pairs ${\bf 2}+\overline{\bf 2}$ could arise from $A_2$ type singularity.  
\par Summarising the above, for F-theory compactification on a Fermat type K3 $\times$ a K3, matter arising from an $E_6$-singularity are only the adjoints {\bf 35} of $SU(6)$ without flux. The vector-like pairs ${\bf 20}+\overline{\bf 20}$ could also arises on 7-branes when flux is turned on. Matter fields arising from an $A_2$-singularity are only the adjoints {\bf 3} of $SU(3)$ without flux, and the vector-like pairs ${\bf 2}+\overline{\bf 2}$ could also arise on 7-branes by including flux.
\par For F-theory compactifications on $S_{[6 \hspace{1mm} 3 \hspace{1mm} 6]}$ $\times$ $S_{[4 \hspace{1mm} 2 \hspace{1mm} 4]}$ and on $S_{[6 \hspace{1mm} 3 \hspace{1mm} 6]}$ $\times$ $S_{[2 \hspace{1mm} 1 \hspace{1mm} 2]}$, the tadpole anomaly is cancelled by including sufficiently many 3-branes. Therefore, for these 2 models the vector-like pairs ${\bf 20}+\overline{\bf 20}$ {\it will} arise by including flux. Recall that $S_{[6 \hspace{1mm} 3 \hspace{1mm} 6]}$ represents a Fermat type K3 surface with $E_6\times E_6\times E_6$ gauge groups. This attractive K3 surface has singular fibers only of type $IV^*$, so all the singularities on $S_{[6 \hspace{1mm} 3 \hspace{1mm} 6]}$ have type $E_6$. It follows that only the pairs ${\bf 20}+\overline{\bf 20}$ arise from the singularities as vector-like pairs. So, the full matter spectrum with flux for F-theory compactifications on $S_{[6 \hspace{1mm} 3 \hspace{1mm} 6]}$ $\times$ $S_{[4 \hspace{1mm} 2 \hspace{1mm} 4]}$ and on $S_{[6 \hspace{1mm} 3 \hspace{1mm} 6]}$ $\times$ $S_{[2 \hspace{1mm} 1 \hspace{1mm} 2]}$ are the adjoints {\bf 35} of $SU(6)$ and the vector-like pairs ${\bf 20}+\overline{\bf 20}$.           

\subsection{Matter Fields for K3 surfaces of Hesse Form (times K3)}
As we saw in section 3, singular fibers of a K3 surface of Hesse form are generically of type $I_3$. For some special members among the whole family of K3 surfaces of Hesse form, two $I_3$ fibers collide and enhance to $I_6$ fiber. The singularity types corresponding to fiber types $I_3$ and $I_6$ are $A_2$ and $A_5$, respectively. 
\par We saw the matter fields which can arise from $A_2$-singularity just the above: the adjoint {\bf 3} of $SU(3)$ and the vector-like pair ${\bf 2}+\overline{\bf 2}$. (In general, we can only say that the vector-like pair ${\bf 2}+\overline{\bf 2}$ could arise by including flux.) It remains to compute the matter fields which can arise from $A_5$-singularity. The enhancement is 
\begin{equation}
A_4\subset A_5.
\end{equation}
The adjoint of $A_5$ decomposes into irreducible representations of $A_4$ as 
\begin{equation}
{\bf 35}={\bf 24}+{\bf 5}+\overline{\bf 5}+{\bf 1},
\end{equation}
where {\bf 35} and {\bf 24} are the adjoints of $SU(6)$ and $SU(5)$ respectively. The pair ${\bf 5}+\overline{\bf 5}$ forms an $N=2$ hypermultiplet. By including flux, the hypermultiplets are split into $N=1$ vector-like pairs ${\bf 5}+\overline{\bf 5}$.  
\par Summarising the above, for F-theory compactification on a {\it generic} K3 surface of Hesse form $\times$ a K3, the only matter on 7-branes are the adjoints {\bf 3} of $SU(3)$ without flux. The vector-like pairs ${\bf 2}+\overline{\bf 2}$ could also arise by including flux. For F-theory on some special K3 surface of Hesse form (with a singular fiber of enhanced type $I_6$) $\times$ a K3, the adjoints {\bf 24} of $SU(5)$ also arise on 7-branes without flux. The vector-like pairs ${\bf 5}+\overline{\bf 5}$ could also arise with flux turned on.     

\section{Conclusion}
We considered F-theory compactified on a product K3 $\times$ K3, with one of the K3's chosen to be of Fermat type or in Hesse form, as introduced in section 2. K3 surfaces of Fermat type and K3 surfaces in Hesse form do not have a section, so such constructed K3 $\times$ K3 does not admit a section to the fibration, as an elliptic Calabi-Yau 4-fold. Therefore, F-theory compactifications on these constructed K3 $\times$ K3 provide models without a section. 
\par We determined gauge groups and matter fields which arise on 7-branes for these F-theory models without a section. For F-theory compactifications on both a Fermat type K3 $\times$ a K3 and a K3 surface in Hesse form $\times$ a K3, gauge groups arising on 7-branes are generically $SU(3)^{\oplus 6}$. When two singular fibers collide, $SU(3)\oplus SU(3)$ enhances to an exceptional gauge group $E_6$ for F-theory on a Fermat type K3 $\times$ a K3, and enhances to $SU(6)$ for F-theory on a K3 surface in Hesse form $\times$ a K3. So exceptional gauge group $E_6$ appears for some of F-theory compactifications on a Fermat type K3 $\times$ a K3. We completely classified the configurations of gauge groups arising on 7-branes, for F-theory compactifications on a Fermat type K3 $\times$ a K3. There are 4 cases in total. K3 $\times$ K3 we consider in this paper does not have a section. So, the gauge groups arising on 7-branes do not have $U(1)$-part.
\par We saw that the gauge groups arising on 7-branes we computed are in agreement with the tadpole cancellation condition. The tadpole cancellation requires that there be 24 7-branes in total, and we confirmed that our solutions meet this consistency condition. We also saw that consideration on monodromy uniquely determines the gauge symmetries for F-theory compactifications on a Fermat type K3 $\times$ a K3. 
\par Since all the 7-branes are parallel for F-theory compactified on K3 $\times$ K3, Yukawa-type interaction is absent for our F-theory models on K3 $\times$ K3. There are no matter curves. F-theory on K3 $\times$ K3 gives 4d theory with $N=2$ supersymmetry. By including flux, half the supersymmetry is broken, left with $N=1$. Only light matter are adjoint matters of gauge groups on 7-branes without flux. When flux is turned on, vector-like pairs could also arise on 7-branes. In general, we can only say that the vector-like pairs could arise, because potential candidate for matter spectrum may actually vanish due to anomalies. For some special cases of F-theory compactifications on a Fermat type K3 $\times$ a K3, we saw that the tadpole anomaly can be cancelled. For these special cases, vector-like pairs actually arise.    

\section*{Acknowledgments}
We would like to thank Tohru Eguchi and Shigeru Mukai for discussions. We are also grateful to the referee for improving this manuscript. This work is supported by Grant-in-Aid for JSPS Fellows No. 26$\cdot$2616.


\newpage
\begin{thebibliography}{99}
\bibitem{Vaf}C.~Vafa, ``Evidence for F-theory'', {\it Nucl. Phys.} {\bf B 469} (1996) 403 [arXiv:hep-th/9602022].
\bibitem{MV1}D.~R.~Morrison and C.~Vafa, ``Compactifications of F-theory on Calabi-Yau threefolds. 1'', {\it Nucl. Phys.} {\bf B 473} (1996) 74 [arXiv:hep-th/9602114].
\bibitem{MV2}D.~R.~Morrison and C.~Vafa, ``Compactifications of F-theory on Calabi-Yau threefolds. 2'', {\it Nucl. Phys.} {\bf B 476} (1996) 437 [arXiv:hep-th/9603161].
\bibitem{Kod}K.~Kodaira, ``On compact analytic surfaces II, III'', {\it Ann. of Math.} {\bf 77} (1963), 563--626; {\it Ann. of Math.} {\bf 78} (1963), 1--40.  

\bibitem{DW}R.~Donagi and M.~Wijnholt, ``Breaking GUT Groups in F-Theory'', {\it Adv. Theor. Math. Phys.} {\bf 15} (2011) 1523--1603 [arXiv:0808.2223 [hep-th]].
\bibitem{FI}A.~Font and L.~E.~Ibanez, ``Yukawa Structure from U(1) Fluxes in F-theory Grand Unification'', {\it JHEP} {\bf 02} (2009) 016 [arXiv:0811.2157 [hep-th]].
\bibitem{CGH}M.~Cveti\v c, I.~Garcia-Etxebarria and J.~Halverson, ``Global F-theory Models: Instantons and Gauge Dynamics'', {\it JHEP} {\bf 01} (2011) 073 [arXiv:1003.5337 [hep-th]].
\bibitem{GW}T.~W.~Grimm and T.~Weigand, ``On Abelian Gauge Symmetries and Proton Decay in Global F-theory GUTs'', {\it Phys. Rev.} {\bf D82} (2010) 086009 [arXiv:1006.0226 [hep-th]].
\bibitem{KMSS}S.~Katz, D.~R.~Morrison, S.~Sch\"afer-Nameki and J.~Sully, ``Tate's algorithm and F-theory'', {\it JHEP} {\bf 08} (2011) 094 [arXiv:1106.3854 [hep-th]].
\bibitem{GKPW}T.~W.~Grimm, M.~Kerstan, E.~Palti and T.~Weigand, ``Massive Abelian Gauge Symmetries and Fluxes in F-theory'', {\it JHEP} {\bf 12} (2011) 004 [arXiv:1107.3842 [hep-th]].
\bibitem{CKP}M.~Cveti\v c, D.~Klevers and H.~Piragua, ``F-Theory Compactifications with Multiple U(1)-Factors: Constructing Elliptic Fibrations with Rational Sections'', {\it JHEP} {\bf 06} (2013) 067 [arXiv:1303.6970 [hep-th]].
\bibitem{MPW}C.~Mayrhofer, E.~Palti and T.~Weigand, ``U(1) symmetries in F-theory GUTs with multiple sections'', {\it JHEP} {\bf 03} (2013) 098 [arXiv:1211.6742 [hep-th]].
\bibitem{BMPW}M.~Bies, C.~Mayrhofer, C.~Pehle and T.~Weigand, ``Chow groups, Deligne cohomology and massless matter in F-theory'', [arXiv:1402.5144 [hep-th]].
\bibitem{CGKP}M.~Cveti\v c, A.~Grassi, D.~Klevers and H.~Piragua, ``Chiral Four-Dimensional F-Theory Compactifications With SU(5) and Multiple U(1)-Factors'', {\it JHEP} {\bf 04} (2014) 010 [arXiv:1306.3987 [hep-th]].
\bibitem{BGK}V.~Braun, T.~W.~Grimm and J.~Keitel, ``New Global F-theory GUTs with U(1) symmetries'', {\it JHEP} {\bf 09} (2013) 154 [arXiv:1302.1854 [hep-th]].

\bibitem{BEFNQ}P.~Berglund, J.~Ellis, A.~E.~Faraggi, D.~V.~Nanopoulos and Z.~Qiu, ``Elevating the free fermion $Z_2\times Z_2$ orbifold model to a compactification of F theory'', {\it Int. Jour. of Mod. Phys.} {\bf A 15} (2000) 1345--1362 [arXiv:hep-th/9812141].
\bibitem{BM}V.~Braun and D.~R.~Morrison, ``F-theory on Genus-One Fibrations'', {\it JHEP} {\bf 08} (2014) 132 [arXiv:1401.7844 [hep-th]].
\bibitem{MT section}D.~R.~Morrison and W.~Taylor, ``Sections, multisections, and $U(1)$ fields in F-theory'', [arXiv:1404.1527 [hep-th]].
\bibitem{AGGK}L.~B.~Anderson, I.~Garcia-Etxebarria, T.~W.~Grimm and J.~Keitel, ``Physics of F-theory compactifications without section'', {\it JHEP} {\bf 12} (2014) 156 [arXiv:1406.5180 [hep-th]].
\bibitem{KMOPR}D.~Klevers, D.~K.~Mayorga Pena, P.~K.~Oehlmann, H.~Piragua and J.~Reuter, ``F-Theory on all Toric Hypersurface Fibrations and its Higgs Branches'', {\it JHEP} {\bf 01} (2015) 142 [arXiv:1408.4808 [hep-th]].
\bibitem{GGK}I.~Garcia-Etxebarria, T.~W.~Grimm and J.~Keitel, ``Yukawas and discrete symmetries in F-theory compactifications without section'', {\it JHEP} {\bf 11} (2014) 125 [arXiv:1408.6448 [hep-th]].
\bibitem{MPTW}C.~Mayrhofer, E.~Palti, O.~Till and T.~Weigand, ``Discrete Gauge Symmetries by Higgsing in four-dimensional F-Theory Compactifications'', {\it JHEP} {\bf 12} (2014) 068 [arXiv:1408.6831 [hep-th]].
\bibitem{MPTW2}C.~Mayrhofer, E.~Palti, O.~Till and T.~Weigand, ``On Discrete Symmetries and Torsion Homology in F-Theory'', {\it JHEP} {\bf 06} (2015) 029 [arXiv:1410.7814 [hep-th]].
\bibitem{CDKPP}M.~Cveti\v c, R.~Donagi, D.~Klevers, H.~Piragua and M.~Poretschkin, ``F-theory vacua with $\mathbb Z_3$ gauge symmetry'', {\it Nucl. Phys.} {\bf B898} (2015) 736--750 [arXiv:1502.06953 [hep-th]].
\bibitem{LMTW}L.~Lin, C.~Mayrhofer, O.~Till and T.~Weigand, ``Fluxes in F-theory Compactifications on Genus-One Fibrations'', [arXiv:1508.00162 [hep-th]].

\bibitem{KV}S.~H.~Katz and C.~Vafa, ``Matter from geometry'', {\it Nucl. Phys.} {\bf B 497} (1997) 146 [arXiv:hep-th/9606086].
\bibitem{BIKMSV}M.~Bershadsky, K.~A.~Intriligator, S.~Kachru, D.~R.~Morrison, V.~Sadov and C.~Vafa, ``Geometric singularities and enhanced gauge symmetries'', {\it Nucl. Phys.} {\bf B 481} (1996) 215 [arXiv:hep-th/9605200].
\bibitem{KMP}S.~H.~Katz, D.~R.~Morrison and M.~R.~Plesser, ``Enhanced gauge symmetry in type II string theory'', {\it Nucl. Phys.} {\bf B 477} (1996) 105 [arXiv:hep-th/9601108].
\bibitem{Pha}E.~Witten, ``Phase transitions in M-theory and F-theory'', {\it Nucl. Phys.} {\bf B 471} (1996) 195, [arXiv:hep-th/9603150].
\bibitem{MT matter}D.~R.~Morrison and W.~Taylor, ``Matter and singularities'', {\it JHEP} {\bf 01} (2012) 022 [arXiv:1106.3563 [hep-th]]. 
\bibitem{RS}G.~V.~Ravindra and V.~Srinivas, ``The Noether-Lefschetz theorem for the divisor class group'', {\it J.Algebra} {\bf 322} (2009), No. 9, 3373--3391.
\bibitem{Ott}J.~C.~Ottem, ``Birational geometry of hypersurfaces in products of projective spaces'', Mathematische Zeitschrift {\bf 280}, Issue 1--2 (2015), 135--148 [arXiv:1305.0537 [math.AG]].
\bibitem{Tate}J.~Tate, ``Algorithm for determining the type of a singular fiber in an elliptic pencil'', in Modular Functions of One Variable IV, Springer, Berlin (1975), 33--52.
\bibitem{SVW}S.~Sethi, C.~Vafa and E.~Witten, ``Constraints on low dimensional string compactifications'', {\it Nucl. Phys.} {\bf B 480} (1996) 213--224, [arXiv: hep-th/9606122].
\bibitem{M}G.~W.~Moore, ``Les Houches lectures on strings and arithmetic'', [arXiv: hep-th/0401049]. 
\bibitem{BKW}A.~P.~Braun, Y.~Kimura and T.~Watari, ``The Noether-Lefschetz problem and gauge-group-resolved landscapes: F-theory on K3 $\times$ K3 as a test case'', {\it JHEP} {\bf 04} (2014) 050 [arXiv:1401.5908 [hep-th]].
\bibitem{SI}T.~Shioda and H.~Inose, ``On Singular K3 surfaces'', in W.~L.~Jr.~Baily and T.~Shioda (eds.), {\it Complex analysis and algebraic geometry}, Iwanami Shoten, Tokyo (1977), 119--136.
\bibitem{AK}P.~S.~Aspinwall and R. Kallosh, ``Fixing all moduli for M-theory on K3$\times$K3'', {\it JHEP} {\bf 10} (2005) 001 [arXiv: hep-th/0506014].
\bibitem{W}E.~Witten, ``On flux quantization in M theory and the effective action'', {\it J. Geom. Phys.} {\bf 22} (1997) 1--13 [arXiv: hep-th/9609122].

\bibitem{BCV}A.~P.~Braun, A.~Collinucci and R.~Valandro, ``G-flux in F-theory and algebraic cycles'', {\it Nucl. Phys.} {\bf B 856} (2012) 129 [arXiv:1107.5337 [hep-th]].
\bibitem{BHV1}C.~Beasley, J.~J.~Heckman and C.~Vafa, ``GUTs and Exceptional Branes in  F-theory -I'', {\it JHEP} {\bf 01} (2009) 058 [arXiv:0802.3391 [hep-th]].
\bibitem{BHV2}C.~Beasley, J.~J.~Heckman and C.~Vafa, ``GUTs and Exceptional Branes in  F-theory -II: Experimental Predictions'', {\it JHEP} {\bf 01} (2009) 059 [arXiv:0806.0102 [hep-th]].
\bibitem{TT}P.~K.~Tripathy and S.~P.~Trivedi, ``Compactification with flux on K3 and tori'', {\it JHEP} {\bf 03} (2003) 028 [arXiv: hep-th/0301139].
\bibitem{ADFL}L.~Andrianopoli, R.~D'Auria, S.~Ferrara and M.~A.~Lledo, ``4-D gauged supergravity analysis of type IIB vacua on K3 $\times$ $T^2$/$\Z_2$'', {\it JHEP} {\bf 03} (2003) 044 [arXiv: hep-th/0302174].
\bibitem{GVW}S.~Gukov, C.~Vafa and E.~Witten, ``CFT's from Calabi-Yau four folds'', {\it Nucl. Phys.} {\bf B584} (2000) 69--108 [arXiv: hep-th/9906070].
\bibitem{DRS}K.~Dasgupta, G.~Rajesh and S.~Sethi, ``M theory, orientifolds and G-flux'', {\it JHEP} {\bf 08} (1999) 023 [arXiv: hep-th/9908088]. 
\bibitem{BS}M.~Bershadsky and V.~Sadov, ``F theory on K3 $\times$ K3 and instantons on 7-branes'', {\it Nucl. Phys.} {\bf B510} (1998) 232--246 [arXiv: hep-th/9703194].
\bibitem{BB}K.~Becker and M.~Becker, ``M theory on eight manifolds'', {\it Nucl. Phys.} {\bf B477} (1996) 155--167 [arXiv: hep-th/9605053].
\bibitem{VW}C.~Vafa and E.~Witten, ``A One loop test of string duality'', {\it Nucl. Phys.} {\bf B447} (1995) 261--270 [arXiv:hep-th/9505053]. 
\bibitem{KM}S.~H.~Katz and D.~R.~Morrison, ``Gorenstein threefold singularities with small resolutions via invariant theory for Weyl groups'', {\it Jour. Alg. Geom.} {\bf 1} (1992) 449 [arXiv:alg-geom/9202002].
\bibitem{Sla}R.~Slansky, ``Group theory for unified model building'', {\it Physics Reports} {\bf 79} (1981) 1--128. 

\bibitem{Sch}M.~Sch\"{u}tt and T.~Shioda, ``Elliptic Surfaces'', in {\it Algebraic Geometry in East Asia (Seoul 2008)}, {\it Advanced Studies in Pure Mathematics} {\bf 60} (2010) 51--160 [arXiv:0907.0298 [math.AG]].
\bibitem{Nish}K.-I.~Nishiyama, ``The Jacobian fibrations on some K3 surfaces and their Mordell-Weil groups'', {\it Japan. J. Math.} {\bf 22} (1996), 293--347.
\bibitem{Shioda}T.~Shioda, ``On the Mordell-Weil lattices'', {\it Comment. Math. Univ. St. Pauli} {\bf 39} (1990), 211--240. 
\bibitem{Mor}D.~R.~Morrison, ``On K3 surfaces with large Picard number'', {\it Invent. Math.} {\bf 75} (1984) 105--121.
\bibitem{Nik}V.~V.~Nikulin, ``Integral symmetric bilinear forms and some of their applications'', {\it Math. USSR Izv.} {\bf 14} (1980) 103--167. 
\bibitem{SM}T.~Shioda and N.~Mitani, ``Singular abelian surfaces and binary quadratic forms'', in {\it Classification of algebraic varieties and compact complex manifolds}, Lecture Notes in Mathematics {\bf 412}, Springer-Verlag (1974), 259--287. 
\bibitem{Nie}H.-V.~Niemeier, ``Definite quadratische Formen der Dimension 24 und Diskriminante 1'', {\it J. Number Theory} {\bf 5}, (1973), 142--178.
\bibitem{Shioda modular}T.~Shioda, ``On elliptic modular surfaces'', {\it J. Math. Soc. Japan} {\bf 24} (1972), 20--59. 
\bibitem{PS-S}I.~I.~Piatetski-Shapiro and I.~R.~Shafarevich, ``Torelli's theorem for algebraic surfaces of type {\rm K3}'', {\it Izv. Akad. Nauk SSSR Ser. Mat.} {\bf 35} (1971), 530--572.
\bibitem{Ner}A.~N\'eron, ``Mod\`eles minimaux des vari\'et\'es ab\'eliennes sur les corps locaux et globaux'', {\it Inst. Hautes \'Etudes Sci. Publ. Math.} {\bf 21} (1964).
\bibitem{BHPV}W.~Barth, K.~Hulek, C.~Peters and A.~Van de Ven, {\it Compact complex surfaces}, second edition, Springer-Verlag, Berlin (2004).
\bibitem{Silv}J.~H.~Silverman, {\it Advanced Topics in the Arithmetic of Elliptic Curves}, Graduate Texts in Mathematics {\bf 151}, Springer-Verlag (1994).
\bibitem{Cas}J.~W.~S.~Cassels, {\it Lectures on Elliptic Curves}, London Math. Society Student Texts {\bf 24}, Cambridge University Press (1991).
\bibitem{Sel}E.~S.~Selmer, ``The diophantine equation $ax^3+by^3+cz^3=0$'', Acta Mathematica, {\bf 85}(1) (1951), 203--362.

\end{thebibliography}
\end{document}